\documentclass[aps,floatfix,preprint,preprintnumbers,nofootinbib,superscriptaddress,natbib]{revtex4}

\usepackage{graphicx,float}
\usepackage{epsfig}		
\usepackage{dcolumn}
\usepackage{bm}
\usepackage{subfigure}

\usepackage[all]{xy}
\usepackage{amsmath,upgreek}
\usepackage{amssymb}

\usepackage{pdfpages}
\usepackage{color}
\usepackage{graphicx,epstopdf}
\usepackage[colorlinks=true,
 linkcolor=red,
 urlcolor=purple,
 citecolor=blue,hyperindex]{hyperref}
\def\0{\mbox{\tiny $0$}}
\def\1{\mbox{\tiny $1$}}
\def\2{\mbox{\tiny $2$}}
\def\3{\mbox{\tiny $3$}}
\def\4{\mbox{\tiny $4$}}
\def\5{\mbox{\tiny $5$}}
\def\6{\mbox{\tiny $6$}}
\def\7{\mbox{\tiny $7$}}
\def\8{\mbox{\tiny $8$}}
\def\9{\mbox{\tiny $9$}}

\def\f14{\mbox{\tiny $\frac{1}{4}$}}

\begin{document}

\title{Quantum prey-predator dynamics: a gaussian ensemble analysis}

\renewcommand{\baselinestretch}{1.2}
\author{A. E. Bernardini}
\email{alexeb@ufscar.br}
\affiliation{~Departamento de F\'{\i}sica, Universidade Federal de S\~ao Carlos, PO Box 676, 13565-905, S\~ao Carlos, SP, Brasil.}
\author{O. Bertolami}
\email{orfeu.bertolami@fc.up.pt}
\altaffiliation[Also at~]{Centro de F\'isica do Porto, Rua do Campo Alegre 687, 4169-007, Porto, Portugal.} 
\affiliation{Departamento de F\'isica e Astronomia, Faculdade de Ci\^{e}ncias da
Universidade do Porto, Rua do Campo Alegre 687, 4169-007, Porto, Portugal.}
\date{\today}

\begin{abstract}
Quantum frameworks for modeling competitive ecological systems and self-organizing structures have been investigated under multiple perspectives yielded by quantum mechanics.
These comprise the description of the phase-space prey-predator competition dynamics in the framework of the Weyl-Wigner quantum mechanics. In this case, from the classical dynamics described by the Lotka-Volterra (LV) Hamiltonian, quantum states convoluted by statistical gaussian ensembles can be analytically evaluated.
Quantum modifications on the patterns of equilibrium and stability of the prey-predator dynamics can then be identified.
These include quantum distortions over the equilibrium point drivers of the LV dynamics which are quantified through the Wigner current fluxes obtained from an onset Hamiltonian background. 
In addition, for gaussian ensembles highly localized around the equilibrium point, stability properties are shown to be affected by emergent topological quantum domains which, in some cases, could lead either to extinction and revival scenarios or to the perpetual coexistence of both prey and predator agents identified as quantum observables in microscopic systems.
Conclusively, quantum and gaussian statistical driving parameters are shown to affect the stability criteria and the time evolution pattern for such microbiological-like communities.
\end{abstract}


\date{\today}
\maketitle

\section{Introduction.}
Quantum-based frameworks \cite{0001,0002} for modeling competitive ecological systems, which not only account for environmental effects \cite{0003}, but also assume the onset hypothesis of a non-linear dynamics for explaining complex and self-organizing hierarchical structures \cite{0004}, have already been investigated in the context of the paradigms of quantum mechanics (QM) \cite{Novo2021BB,Novo2021}. These encompass some relevant phenomenological issues related, for instance, to prey-predator population oscillations, competition-induced chaos, and molecular programming strategies for symbiotic synchronization \cite{PP00,PP01,PP02,PP03,PP04}. 
Furthermore, from pure biochemical and biological evolutionary perspectives, it is a theoretical possibility that living organisms or biochemical systems exhibit operative mechanisms of quantum mechanical origin, which can be evaluated and statistically interpreted.

From a non-deterministic outlook, the prediction of existence of competitive species in analogy with interacting quantum states \cite{Bio17}, can be regarded as a measurement operation. On its hand, this can be related to a quantum statistical ensemble description, for instance, through a gaussian (single- or multi-particle) phase-space probability distribution.
The quantum statistical distribution interpretation of the ecosystem components may follow the statistics of gaussian ensembles which, in a classical approach, would be identically the same.
This suggests that collective behaviors depicted from phase-space semiclassical trajectories can be the averaging results from the statistical treatment of the space-time evolution of the species distribution densities, $y$ and $z$.
In this case, the quantum observables are identified by canonically conjugate (dimensionless) operators, $\hat{x}$ and $\hat{k}$, such that their averaged out values, $x=\langle\hat{x}\rangle$ and $k=\langle\hat{k}\rangle$, are in correspondence with the species distribution densities by $y= e^{-x}$ and $z = e^{-k}$.
As mentioned, this encompasses a measurement operation for which, even macroscopically, a ground {\em quantum analog} non-commutative property of the phase-space coordinates $\hat{x}$ and $\hat{k}$, $[\hat{x},\,\hat{k}]\neq 0$, is assumed.
In fact, the information incompleteness associated with the probabilistic nature of the quantum-like non-commutative hypothesis, $[\hat{x},\,\hat{k}]\neq 0$, requires a complementary statistical view of the dynamically involved variables, $x$ and $k$.
In particular, in case of $[\hat{x},\,\hat{k}]= i$, with the Planck's constant, $\hbar$, being set equal to unity, classical and quantum dynamics can then be assumed to coexist at different scales, the classical macroscopic one and the quantum microscopic one. 

Through the framework proposed here, the bridge between classical and quantum descriptions for the problem of population oscillation dynamics can be obtained by the Hamiltonian equations of motion when the representation of the statistical distributions for an ensemble of particles is given in terms of gaussian Wigner statistical distributions.
Furthermore, with the phase-space Weyl-Wigner (WW) framework \cite{Wigner,Ballentine,Case} encompassing all the QM paradigms, classical and quantum patterns evaluated for species distributions can be interpreted from the record of their observed occurrence in a given phase-space time-dependent map \cite{Novo2021BB}.

In this context, the Lotka-Volterra (LV) dynamical equations for prey-predator systems \cite{LV1,LV2} are considered as a consistent departing platform for obtaining quantum mechanical modifications.
Besides yielding remarkable results in describing the behavior of macroscopic ecosystems \cite{PRE-LV,SciRep02,PRE-LV2,RPSA-LV,Anna}, the LV system has been considered in a wide range of microscopic scenarios which include, for instance, the description of stability criteria for microbiological communities \cite{Nature01, Nature02}, the emergence of phase transitions in finite microscopic systems \cite {PRE-LV3}, and the support for the dynamically driven stochastic systems \cite{Allen,Grasman}.
Furthermore, classical LV equations have a simple nonlinear Hamiltonian pattern described by the Hamiltonian,
\begin{equation}\label{altHam}
\mathcal{H}(x,\,k) = a \,x + k + a\, e^{-x} + e^{-k},
\end{equation}
with $a > 0$, which results in classical equations of motion resumed by
\begin{eqnarray}
\label{altHam2B}
d{x}/d\tau &=& \{x,\mathcal{H}\}_{PB} = 1-e^{-k},\\
\label{altHam2C}
d{k}/d\tau &=& \{k,\mathcal{H}\}_{PB} = a\,e^{-x} - a,
\end{eqnarray}
for which the condition $\partial^2 \mathcal{H} / \partial x \, \partial k = 0$ is clearly satisfied.
Eqs.~\eqref{altHam2B} and \eqref{altHam2C} drive $x$ and $k$ oscillations correlated to the number of prey and predator species, $y$ and $z$, by $y = e^{-x}$ and $z = e^{-k}$, leading to ecological coexistence chains theoretically depicted by phase-space closed orbits, $\mathcal{H}(x,\,k) = \epsilon$ (cf. Fig.~1 from Ref. \cite{Novo2021BB}), with $\epsilon \in (a+1,\infty)$.

Quantum mechanical features resulting from the Wigner flow of probabilities for generic $1$-dim Hamiltonian systems described by Eq.~(\ref{altHam}) have already been identified and quantified by means of the so-called Wigner currents \cite{Novo2021}.
Stationarity and Liovillianity quantifiers \cite{Zurek02,Steuernagel3,NossoPaper,Meu2018} were also obtained through probability and information fluxes driven by the LV background dynamics \cite{Novo2021,Novo2021BB}. 
This description allowed, for instance, the understanding of the non-Liouvillian and non-stationary issues related to the time evolution of the prey-predator dynamics \cite{Novo2021BB,Novo2023}.
In addition, equilibrium and stability features, since from continuous (hyperbolic regime) up to discrete (chaotic regime) domains, were identified in order to quantify the influence of quantum fluctuations over equilibrium and stability scenarios of LV driven systems \cite{Novo2023}. 

However, all the above-mentioned analysis misses the resulting description of how the time evolving number of species, $y$ and $z$, identified through the first moment of canonical variables, $x$ and $k$, are related to the averaged quantum driven phase-space trajectories, and to the stability and quantum distortions over the equilibrium point directives.
Once quantified through the Wigner current fluxes, a semiclassical interpretation of $(x,\,k) =(\langle\hat{x}\rangle,\,\langle\hat{k}\rangle)$, which are then computed under the influence of quantum fluctuations (i.e. with $[\hat{x},\,\hat{k}]= i$), can be indeed encompassed by the generalized WW phase-space framework \cite{Novo2021}.
Hence, in case of highly localized gaussian ensembles around the equilibrium point which shall be here discussed, non-extinction, discreteness and stability properties associated with emergent topological quantum domains can be shown to be all closely connected. 
More relevantly, they will provide the quantum modified phase-space scenario from which semiclassical trajectories describing time-evolution of competing species are identified.

Thus, the outline of the manuscript is as follows.
Methods for obtaining quantum driven phase-space trajectories \cite{Novo2021} in the Wigner flow are reported in Sec. II.
Analytical results are specialized to gaussian ensembles driven by the LV Hamiltonian, Eq.~(\ref{altHam}), in order to account for the overall quantum distortion pattern over the classical regime \cite{Novo2021BB}.
The quantitative results are presented in Sec. III. They include the description of the quantum mechanical drivers of the equilibrium point as well as their effects on the quantum modified time evolution of the number of species.
Conclusions are presented in Sec. IV, and suggest that, in case of microscopic systems for which the quantum approach is relevant, the evinced instability patterns on the prey-predator phase-space trajectories can be interpreted as an output of quantum origin.

\section{Quantum driven phase-space trajectories in the Wigner flow framework}

The Wigner function is identified with a quantum mechanical density matrix operator, $\hat{\rho} = |\psi \rangle \langle \psi |$, through its Weyl transform written in a dimensionless form ($\hbar \sim 1$) as 
{\footnote{Akin to the statistical QM, the WW phase-space framework recovers the QM probabilistic interpretation through its marginal distributions that correspond to position and momentum probability densities given by
\begin{equation}
\vert \psi_x(x)\vert^2 = \int^{+\infty}_{-\infty} \hspace{-.35cm}dx\,\mathcal{W}(x,\, k)
\leftrightarrow
\vert \psi_k(k)\vert^2 = \int^{+\infty}_{-\infty} \hspace{-.35cm}dk\,\mathcal{W}(x,\, k),
\quad\mbox{with}\quad
\psi_x(x) =
\frac{1}{2\pi}\int^{+\infty}_{-\infty} \hspace{-.35cm} dk\,\exp{\left[-i \, k \,x\right]}\,
\psi_k(k),
\end{equation}
which, at $1$-dim, are straightforwardly encompassed by the Heisenberg-Weyl algebra in the form of a position-momentum non-commutative relation, $[x,\,k] = i$.}
\begin{equation}
\hat{\rho} \to \mathcal{W}(x,\, k) = \pi^{-1} 
\int^{+\infty}_{-\infty} \hspace{-.35cm}dy\,\exp{\left[2\, i \, k \,y/\hbar\right]}\,
\psi(x - y)\,\psi^{\ast}(x + y).\label{alt222}
\end{equation}

From elementary properties of \eqref{alt222} \cite{Case,Ballentine}, information and probability fluxes \cite{Steuernagel3,NossoPaper,Meu2018} can be derived from the dimensionless continuity equation stated as 
\begin{equation}\label{altz51dim}
{\partial_{\tau} \mathcal{W}} + \mbox{\boldmath $\nabla$}_{\xi}\cdot\mbox{\boldmath $\mathcal{J}$}={\partial_{\tau} \mathcal{W}} + {\partial_x \mathcal{J}_x}+{\partial_k \mathcal{J}_k}  =0,
\end{equation}
where $\tau$ is the dimensionless time variable and, for an unrestricted Hamiltonian dynamics driven by $\mathcal{H}(x,\,k) = \mathcal{K}(k) + \mathcal{V}(x)$, the vector flow connected to $\mathcal{W}$ is identified by the Wigner currents \cite{Novo2021},
\begin{eqnarray}
\label{altimWAmm}\mathcal{J}_x(x, \, k;\,\tau) &=& +\sum_{\eta=0}^{\infty} \left(\frac{i}{2}\right)^{2\eta}\frac{1}{(2\eta+1)!} \, \left[\partial_k^{2\eta+1}\mathcal{K}(k)\right]\,\partial_x^{2\eta}\mathcal{W}(x, \, k;\,\tau),\\
\label{altimWBmm}\mathcal{J}_k(x, \, k;\,\tau) &=& -\sum_{\eta=0}^{\infty} \left(\frac{i}{2}\right)^{2\eta}\frac{1}{(2\eta+1)!} \, \left[\partial_x^{2\eta+1}\mathcal{V}(x)\right]\,\partial_k^{2\eta}\mathcal{W}(x, \, k;\,\tau).
\end{eqnarray}
The Wigner currents evince the quantum back reaction, which is driven by $\eta > 1$ contributions\footnote{Related to the powers of $\hbar$, $\hbar^{2\eta}$, suppressed from the dimensionless notation.} in the above series expansion, from which the classical Liouvillian regime \cite{Case,Ballentine} limit is identified by the $\eta = 0$ contribution.

Quantum effects on the classical regime are quantified by stationarity and Liouvillianity vector field divergence quantifiers, $\mbox{\boldmath $\nabla$}_{\xi}\cdot\mbox{\boldmath $\mathcal{J}$}$ and $\mbox{\boldmath $\nabla$}_{\xi} \cdot \mathbf{w}$, 
where $\mbox{\boldmath $\xi$} = (\xi_x,\,\xi_k)\equiv (x,\,k)$.
In spite of being independent one from each other, $\mbox{\boldmath $\nabla$}_{\xi}\cdot\mbox{\boldmath $\mathcal{J}$}$ and $\mbox{\boldmath $\nabla$}_{\xi} \cdot \mathbf{w}$ can be mutually connected by constraining $\mbox{\boldmath $\mathcal{J}$}$ and $\mathcal{W}$, in terms of $\mathbf{w} = \mbox{\boldmath $\mathcal{J}$}/\mathcal{W}$, i.e., the {\em quantum analog} of the classical velocity $\mathbf{v}_{\xi(\mathcal{C})} = \dot{\mbox{\boldmath $\xi$}} = (\dot{x},\,\dot{k})\equiv ({\partial_k \mathcal{H}},\,-{\partial_x \mathcal{H}})$. They are given by
\begin{equation} \label{althelps}
\mbox{\boldmath $\nabla$}_{\xi}\cdot\mbox{\boldmath $\mathcal{J}$} = \sum_{\eta=0}^{\infty}\frac{(-1)^{\eta}}{2^{2\eta}(2\eta+1)!} \, \left\{
\left[\partial_x^{2\eta+1}\mathcal{V}(x)\right]\,\partial_k^{2\eta+1}\mathcal{W}
-
\left[\partial_k^{2\eta+1}\mathcal{K}(k)\right]\,\partial_x^{2\eta+1}\mathcal{W}
\right\},\end{equation}
whose vanishing means stationarity, and  
\begin{equation}\label{altdiv2}
\mbox{\boldmath $\nabla$}_{\xi} \cdot \mathbf{w} = \sum_{\eta=0}^{\infty}\frac{(-1)^{\eta}}{2^{2\eta}(2\eta+1)!}
\left\{
\left[\partial_k^{2\eta+1}\mathcal{K}(k)\right]\,
\partial_x\left[\frac{1}{\mathcal{W}}\partial_x^{2\eta}\mathcal{W}\right]
-
\left[\partial_x^{2\eta+1}\mathcal{V}(x)\right]\,
\partial_k\left[\frac{1}{\mathcal{W}}\partial_k^{2\eta}\mathcal{W}\right]
\right\}, ~~~\end{equation}
whose vanishing means Liouvillianity.

Divergenceless scenarios, with either $\mbox{\boldmath $\nabla$}_{\xi}\cdot\mbox{\boldmath $\mathcal{J}$}=0$ or $\mbox{\boldmath $\nabla$}_{\xi} \cdot \mathbf{w}=0$, allow for recovering local stability conditions (cf. $\partial_{\tau} \mathcal{W} = 0$) and information for distinguishing quantum from Liouvillian regimes (cf. $\mathbf{w} \leftrightarrow \mathbf{v}_{\xi(\mathcal{C})}$), respectively.

Dynamically, for the classical configuration described by the series expansions, Eqs.~\eqref{altimWAmm} and \eqref{altimWBmm}, truncated at $\eta =0$, $x$ and $k$ are constrained by $[\hat{x},\,\hat{k}] = 0$, which means that simultaneous measurements of $x$ and $k$ do not affect each other.
In this case, for the evolution of ensembles of species parameterized by $y = e^{-x}$ and $z = e^{-k}$, the fluctuations of the number of species, parameterized by second-order coordinate momenta, $\delta s = \sqrt{\langle \hat{s}^2\rangle-\langle \hat{s}\rangle^2}$, with $\hat{s} = \hat{x},\, \hat{k}$, follow a deterministic evolution pattern parameterized by the classical velocity, $\mathbf{v}_{\xi(\mathcal{C})}$, Eqs.\eqref{altHam2B} and \eqref{altHam2C}, with $\delta x\,\delta k = 0$.

The {\em quantum analog} hypothesis instead introduces a non-commutative onset constraint, $[\hat{x},\,\hat{k}] = i$, which defines a minimal phase-space elementary cell volume, $\delta x \, \delta k \sim 1$, which can be admitted as a {\em quantum-origin} non-extinction hypothesis.
In this case, $\hat{x}$ and $\hat{k}$ measurements (as well as the corresponding averaged-out species densities, $y$ and $z$) are affected by each other and, quantum mechanically, they cannot be evaluated simultaneously. One then has the {\em quantum analog} of the uncertainty principle expressed by $\delta x\,\delta k \gtrsim 1$, associated to $[\hat{x},\,\hat{k}] = i$.
Since $\delta x\,\delta k \neq 0$, hence $\delta y\,\delta z \neq 0$, there is no deterministic species evolution and, therefore, the fluctuations of the number of species cannot be parameterized, neither by $y(\tau)$ and $z(\tau)$, nor by $y(z)\leftrightarrow z(y)$. In this picture, the Wigner currents drive the statistical and probability effects over the quantum ensembles, yielding semiclassical trajectories. In case of gaussian associated Wigner currents, as it shall be explicitly obtained in the following, the non-linear Hamiltonian components account for the complete contributions from the perturbative expansion, cf. Eqs.~\eqref{altimWAmm} and \eqref{altimWBmm}, which give back the exact expression for the quantum fluctuations \cite{Novo2021,Novo2021BB}.
Effectively, the semiclassical analysis can be performed by replacing the classical Hamiltonian system, Eqs.~\eqref{altHam2B} and \eqref{altHam2C}, by the quantum solution for $\mathbf{w}$ corresponding to the integrated versions of Eqs.~\eqref{altimWAmm} and \eqref{altimWBmm}.

By examining classical to quantum transitions from a statistical viewpoint, the classical analog of the second-order moments of position and momentum coordinates, which parametrize the Heisenberg's uncertainty principle, should consistently satisfy 
the same constraints of position and momentum quantum observables.
It can be shown that such an aasumption is isomorphic to the implementation of a so-called gaussian quantum mechanics \cite{Ballentine,PRAPRAPRA} in the sense that it generates the same statistics.
Therefore, the replacement of the Wigner distributions that appear into Eqs.~\eqref{altimWAmm} and \eqref{altimWBmm} by gaussian distributions given by
\begin{equation}
\mathcal{G}_\alpha(x,\,k) = \frac{\alpha^2}{\pi}\, \exp\left[-\alpha^2\left(x^2+ k^2\right)\right],
\end{equation}
corresponds to the most natural procedure for identifying the classical to quantum transition of Hamiltonian systems which follow an approximated phase-space normal distribution in $x$ and $k$ coordinates\footnote{As previously pointed out, a similar analysis can be carried out for the thermodynamic statistical ensemble interpretation up to $\mathcal{O}(\hbar^2)$ in the series expansion, Eqs.~\eqref{altimWAmm} and \eqref{altimWBmm}.}.
Noticing that gaussian derivatives can be replaced into Eqs.~\eqref{altimWAmm} and \eqref{altimWBmm} by
\begin{equation}\label{altssae}
\partial_\chi^{2\eta+1}\mathcal{G}_{\alpha}(x, \, k) = (-1)^{2\eta+1}\alpha^{2\eta+1}\,\mbox{\sc{h}}_{2\eta+1} (\alpha \chi)\, \mathcal{G}_{\alpha}(x, \, k),\qquad\mbox{for $\chi = x,\, k$,}
\end{equation}
where $\mbox{\sc{h}}_n$ are the Hermite polynomials of order $n$, 
results that gaussian convoluted Wigner currents can be described by well-behaved functions which account for the overall quantum effects over the classical profile \cite{Novo2021}.
Considering the result from Eq.~\eqref{altssae}, and that auxiliary derivatives can be computed from Eq.~\eqref{altHam},
\begin{eqnarray}
\label{altt111B}
\partial_x^{2\eta+1}\mathcal{H}(x,\,k) &=& \delta_{\eta 0} - e^{-k},\\
\label{altt222B}
\partial_k^{2\eta+1}\mathcal{H}(x,\,k) &=& a \left(\delta_{\eta 0} - e^{-x}\right), 
\end{eqnarray}
the gaussian convoluted prey-predator dynamics, Eqs.~\eqref{altimWAmm} and \eqref{altimWBmm}, results\footnote{After noticing that 
\begin{equation}
\sum_{\eta=0}^{\infty}\mbox{\sc{h}}_{2\eta+1} (\alpha \chi)\frac{s^{2\eta+1}}{(2\eta+1)!} = \sinh(2s\,\alpha\chi) \exp[-s^2].
\end{equation}} into the following probability flow contributions \cite{Novo2021BB},
\begin{eqnarray}
\label{altimWA4CC3mm}
\partial_x\mathcal{J}^{\alpha}_x(x, \, k) &=& -2 \left[\alpha^2 \,x - 
\sin\left(\alpha^2\,x\right)\,e^{\frac{\alpha^2}{4}-k}
\right]
\mathcal{G}_{\alpha}(x, \, k),\\
\label{altimWB4CC3mm}
\partial_k\mathcal{J}^{\alpha}_k(x, \, k) &=& +2a\left[\alpha^2\,k - 
\sin\left(\alpha^2\,k\right)\,e^{\frac{\alpha^2}{4}-x}
\right]
\mathcal{G}_{\alpha}(x, \, k),
\end{eqnarray}
from which the corresponding quantum analog Wigner velocities can be expressed by\footnote{In terms of gaussian error functions, $\mbox{\sc{Erf}}[\dots]$.}
\begin{eqnarray}
\label{altimWA4CCD4mm}w^{\alpha}_x(x, \, k) &=& 
1
-\frac{i\,\sqrt{\pi}}{2\alpha} \,e^{-(k-\alpha^2 x^2)}
\left\{\mbox{\sc{Erf}}\left[\alpha(x-i/2)\right]-\mbox{\sc{Erf}}\left[\alpha(x+i/2)\right]\right\},\\
\label{altimWB4CCD4mm}w^{\alpha}_k(x, \, k) &=& 
-a\,\left\{1-
\frac{i\,\sqrt{\pi}}{2\alpha} \,e^{-(x-\alpha^2 k^2)}
\left\{\mbox{\sc{Erf}}\left[\alpha(k-i/2)\right]-\mbox{\sc{Erf}}\left[\alpha(k+i/2)\right]\right\}\right\},\,\,\,\,
\end{eqnarray}
which yield the averaged-out quantum trajectories driven by gaussian distributions, as well as 
quantify stability and quantumness in terms of the divergence quantifiers from Eqs.~\eqref{althelps} and \eqref{altdiv2}, respectively \cite{Novo2023}.

\section{Competing species evolution and instability issues with quantum origin}

From the above results, the Wigner flow pattern can be constrained by the form of $\mbox{\boldmath $\mathcal{J}^{\alpha}$} = \mathbf{w}\,\mathcal{G}_{\alpha}(x, \, k)$ such that gaussianized phase-space prey-predator flow pattern evaluated in terms of the parameter, $\alpha$, can now be interpreted in terms of the phase-space evolution of the equilibrium points identified by $\mathcal{J}^{\alpha}_x = \mathcal{J}^{\alpha}_k=0$, function of $\alpha$, as depicted in Fig.~\ref{altBio02}. The equilibrium points, depicted by their flux surrounding envelops, with boundaries given by $\vert \mathbf{w}\vert < 0.07$ are viewed through different angles. The {\em quasi} stable equilibrium point short displacement is evinced by the blue region, for $\alpha^2 \lesssim 1$. Despite approaching classical-like closed orbits, they are perturbed by a quantum vortex distortion which emerges from surrounding values of $x$ and $k$ that breakdown the equilibrium point stability (lighter white patterns). Naturally, it contributes to the subsequent diffusive appearance of unstable vortices and saddle points that destroy the classical pattern, as evinced by red bubble regions for $\alpha \gtrsim 1$. These correspond to the topological phases \cite{Novo2023} which induce macroscopical modifications onto the prey-predator oscillation pattern. 
\begin{figure}
\vspace{-1.2 cm}
\includegraphics[scale=0.19]{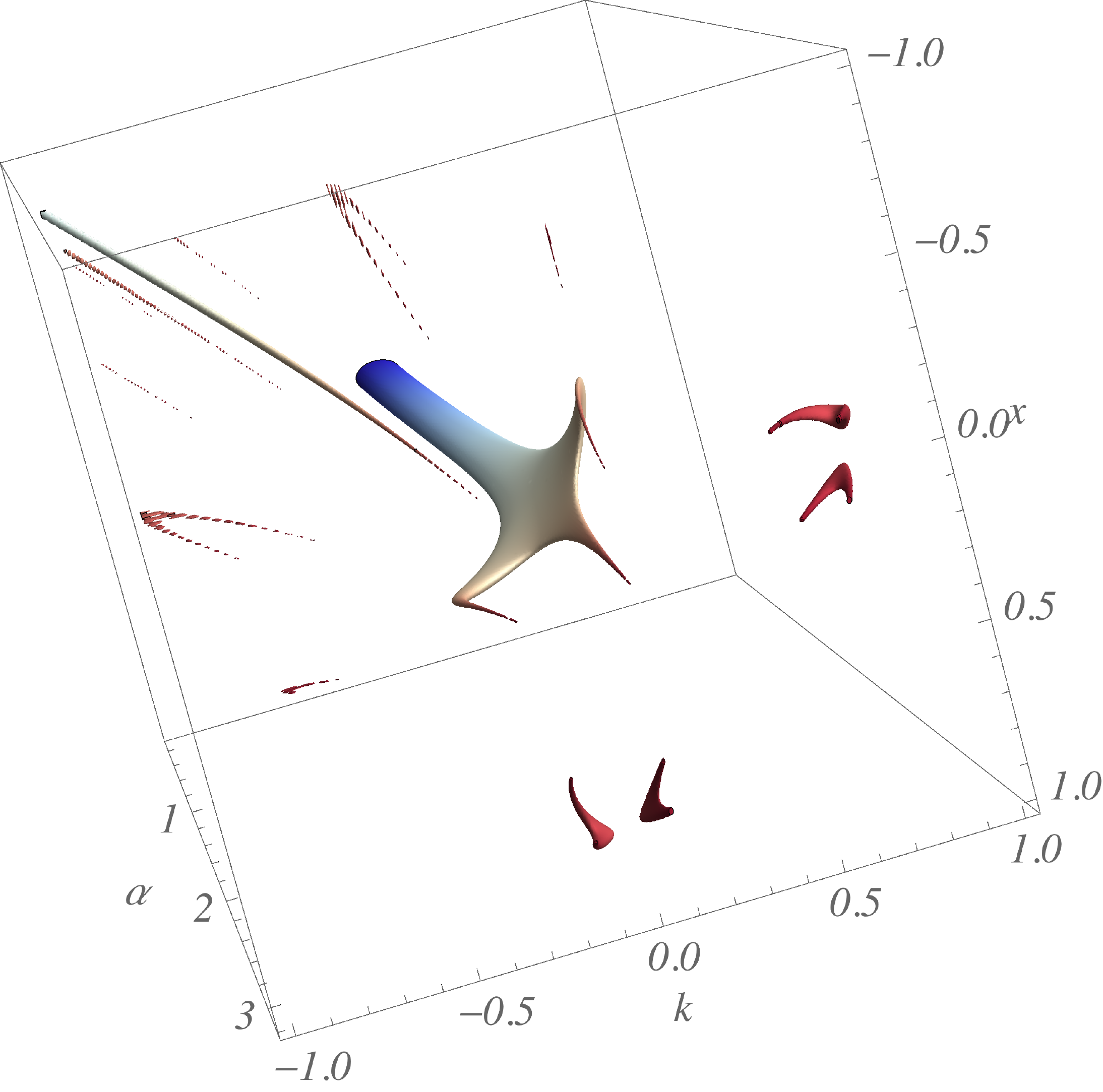}
\includegraphics[scale=0.19]{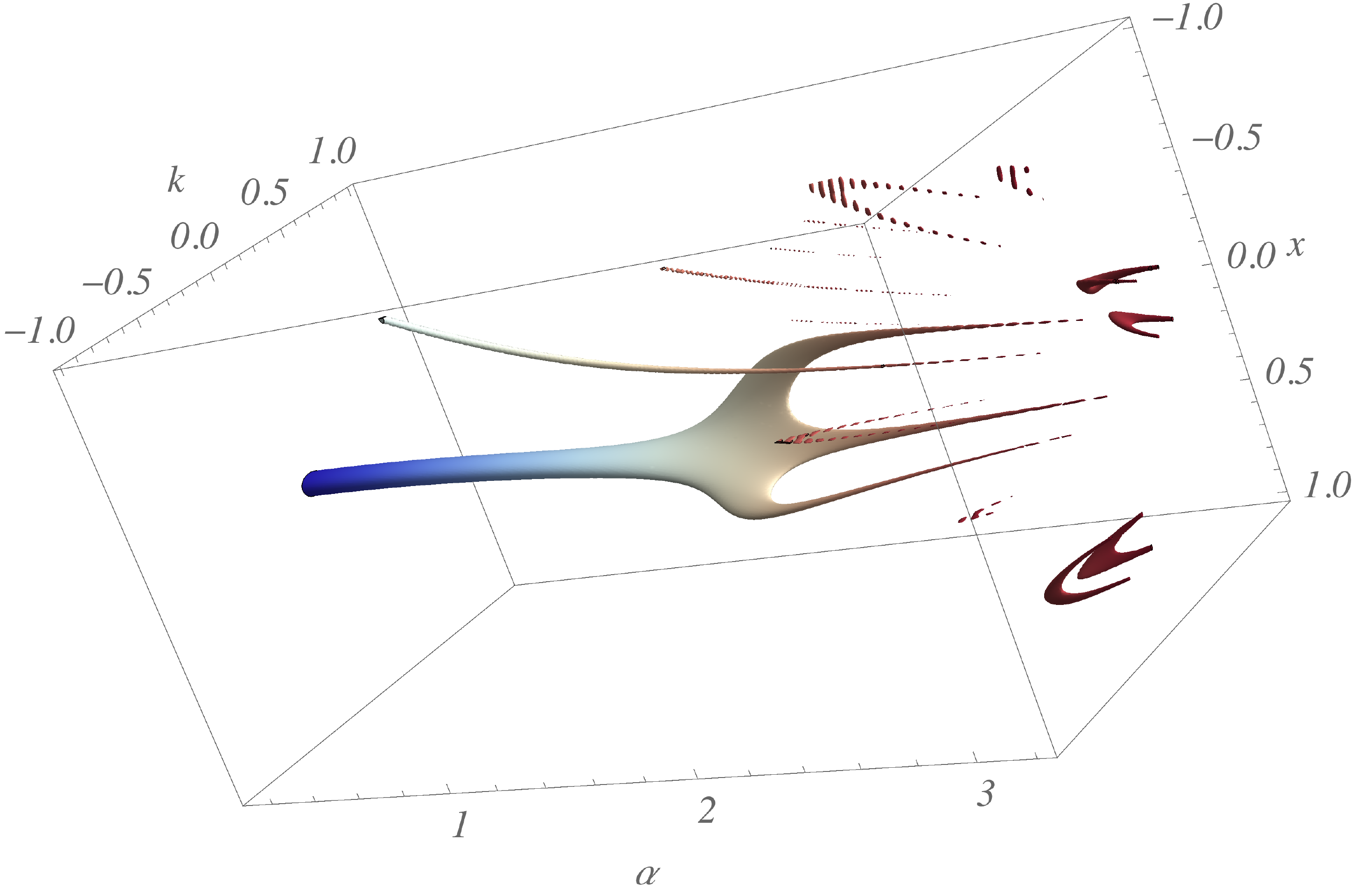}
\includegraphics[scale=0.19]{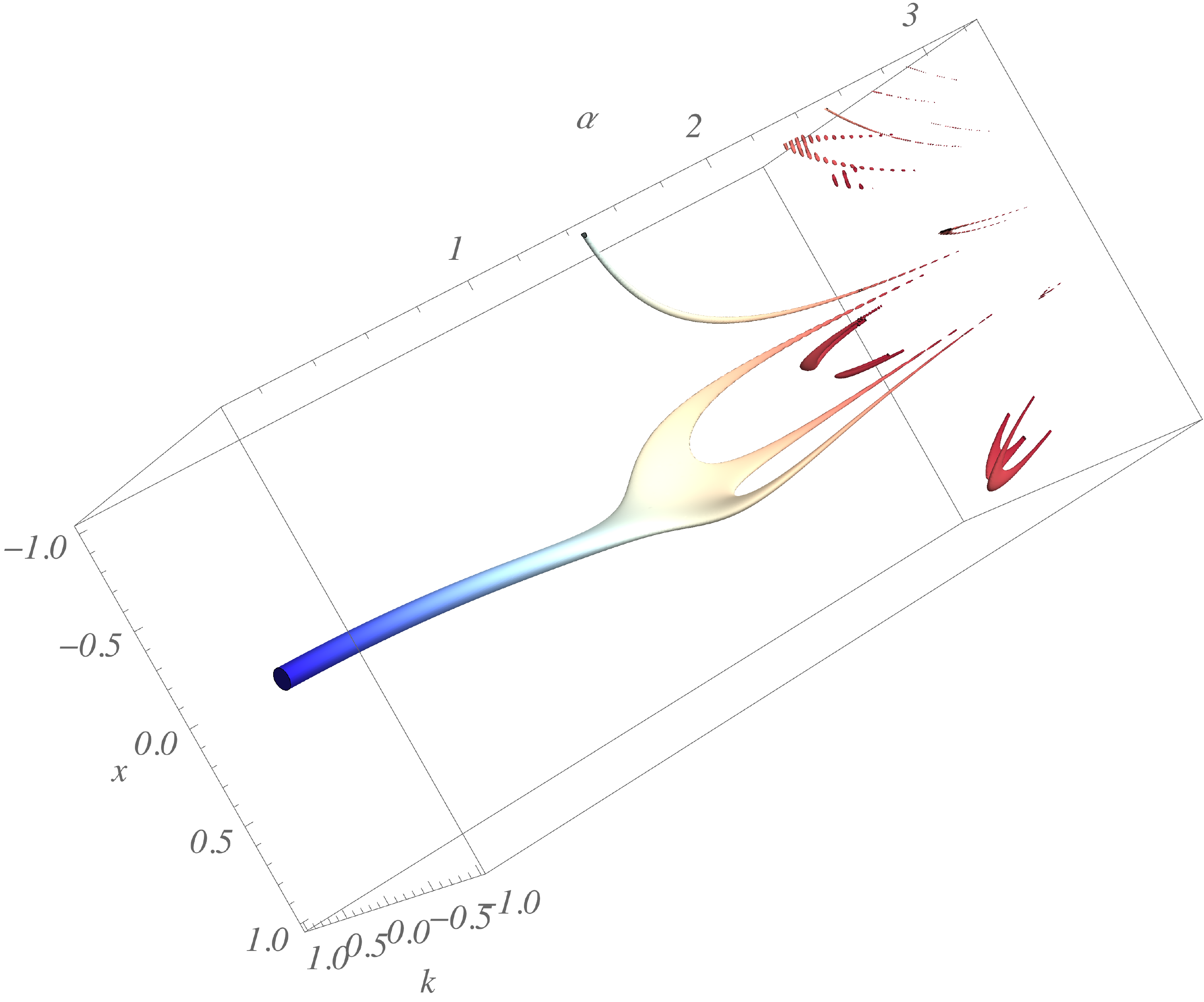}
\renewcommand{\baselinestretch}{.6}
\caption{\footnotesize
(Color online)
Region plot scheme for the phase-space evolution of quantum critical points corresponding to {\em quasi} stable (blue regions) and unstable (white to red regions) equilibrium points in terms of the gaussian spreading $\alpha$.
Results are for the Wigner flow with the equilibrium point (flux) surrounding envelop described by $ \vert\mathbf{w}\vert < 0.07$. Local effects compensate each other when sliced views of the Wigner flux for fixed $\alpha$ are considered, i.e. either when two vortices of opposite winding numbers match each other or when saddle points mutually annihilates one each other.
The spreading behavior of the gaussian ensemble, from red bubble (unstable) islands to the blue ({\em quasi} stable) envelop, corresponding to decreasing values of $\alpha$, diffusively recovers the classical-like pattern for which the quantum imprint is just to the small displacement of the ({\em quasi}) stable equilibrium point. The portraits are the same for different angle views.\label{altBio02}}
\end{figure}

For describing the species evolution and to evaluate the quantum effects on them, one can follow a more specialized analysis of the quantum affected stable patterns through a semiclassical approach which accounts for the distortions obtained from Eqs.~\eqref{altimWA4CC3mm} and \eqref{altimWB4CC3mm}.
By considering the quantum analogue phase-space velocity components from Eqs.~\eqref{altimWA4CCD4mm} and \eqref{altimWB4CCD4mm}, one can numerically obtain the time dependence of prey and predator species, $y(\tau)$ and $z(\tau)$, and see how they are affected by the equilibrium point averaged quantum displacement at almost stable regimes. The results are depicted in Fig.~\ref{altBio03} for typical spreading gaussian ensembles, with $\alpha = 1/4$ and with $a = 1/4,\,1$ and $4$.
As it can be seen, the deviations provided by $a\neq 1$ just indicate positive ($a < 1$) and negative ($a > 1$) fluctuations in the results for the number of species at increasing times. In the former case, prey-predator oscillations are amplified at the same time that the averaged-out quantum effects asymptotically drive the system to periodic extinctions and revivals of the $y(\tau)$ population. Yellow shadows in the first row of Fig.~\ref{altBio03} correspond to an arbitrary quantitative extinction threshold identified by $y(\tau)< 0.04$.
For increasing values of the gaussian $\alpha$-parameter, the result will approach the one from Ref.~\cite{PRE-LV} where discreteness of the populations destroy the mean-field stability and eventually drive the system toward similar effects \cite{Novo2021BB}.
In the latter case, population oscillations are suppressed, also denoting unstable equilibrium.
Again, increasing value of $\alpha$ asymptotically suppresses the oscillation patter around the equilibrium points, which approaches the so-called {\it extinction of oscillating populations} from Ref.~\cite{PRE-LV2}: if $z$ and $y$ are read as prey-predator densities, for highly increasing values of $a$, the non-extinction stable configuration is rapidly reached at $(x,\,k) = (0,0)$, which correspond to $y$ and $z$ equals to unity. 
For $a = 1$, a stable equilibrium point configuration is quickly restored\footnote{In fact, as it can be numerically verified for any gaussian configuration with $\alpha < 1$.}, with the quantum distortions just resulting into a short de-phasing of the population oscillation pattern for longer times, when compared with the classical prediction. 
\begin{figure}
\includegraphics[scale=0.365]{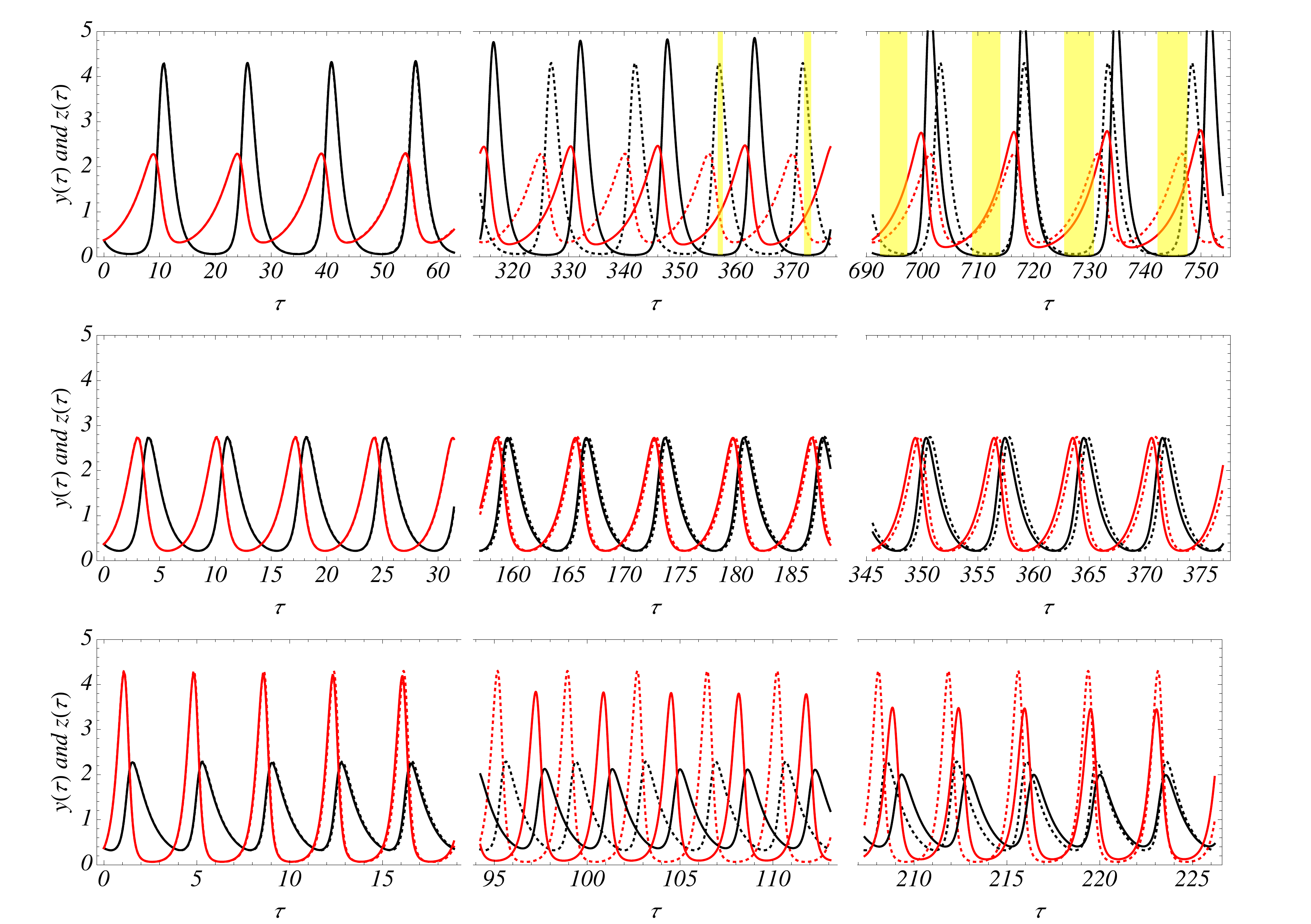}\hspace{-.7cm}
\includegraphics[scale=0.244]{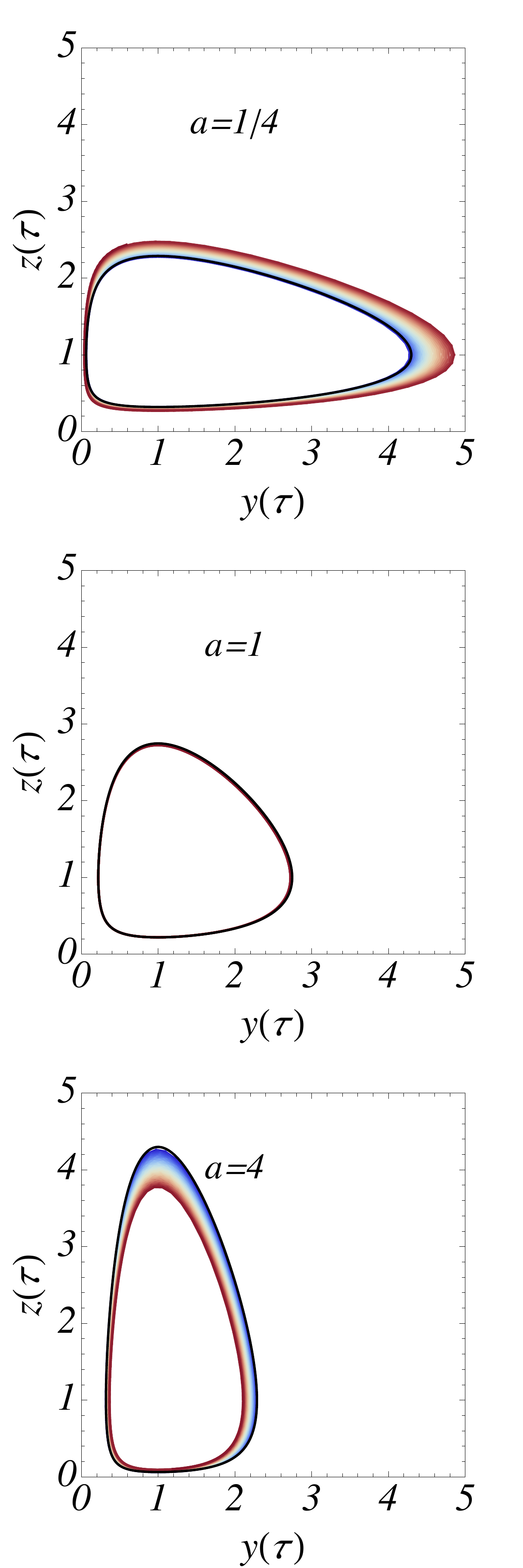}
\renewcommand{\baselinestretch}{.6}
\caption{\footnotesize
(Color online)
{\em First column}: Classical (dashed lines) and quantum (solid lines) prey-predator time-evolution pattern, $y(\tau)$ (red lines) and $z(\tau)$ for typical spreading gaussian ensembles, with $\alpha = 1/4$ and with of $a = 1/4,\,1$ and $4$.
{\em Second column}: Corresponding phase-space ({\em quasi}) stable trajectories for classical (dashed lines) and quantum (solid lines) prey-predator patterns. The color scheme describes the quantum ({\em quasi}) stable evolution from $\tau = 0$ (blue tone) to $\tau \gg 0$ (red tone).
.}
\label{altBio03}
\end{figure}

Nevertheless, the main content of the above depicted results concerns the implications of the quantum hypothesis associated with a statistical ensemble description, which affects significantly the stability pattern.
The inclusion of a statistical distribution gaussian envelope opens the window for quantifying the quantum effects over the prey-predator density pattern in the phase space. An equivalent constant envelope would suppress the quantum effects and recover the classical pattern. Either a gaussian envelope or another phase-space statistical distribution allows for identifying quantum disturbances over the equilibrium points. Conversely, for some modified dynamics, the quantum effects and the prey-predator parameters work, in some sense, as fine-tuned agents that could convert unstable trajectories into stable ones. If the extinction hypothesis is realized as a positive response, one notices that increasing times narrow the revival windows, as depicted in Fig.~\ref{altBio04}.
\begin{figure}
\includegraphics[scale=0.41]{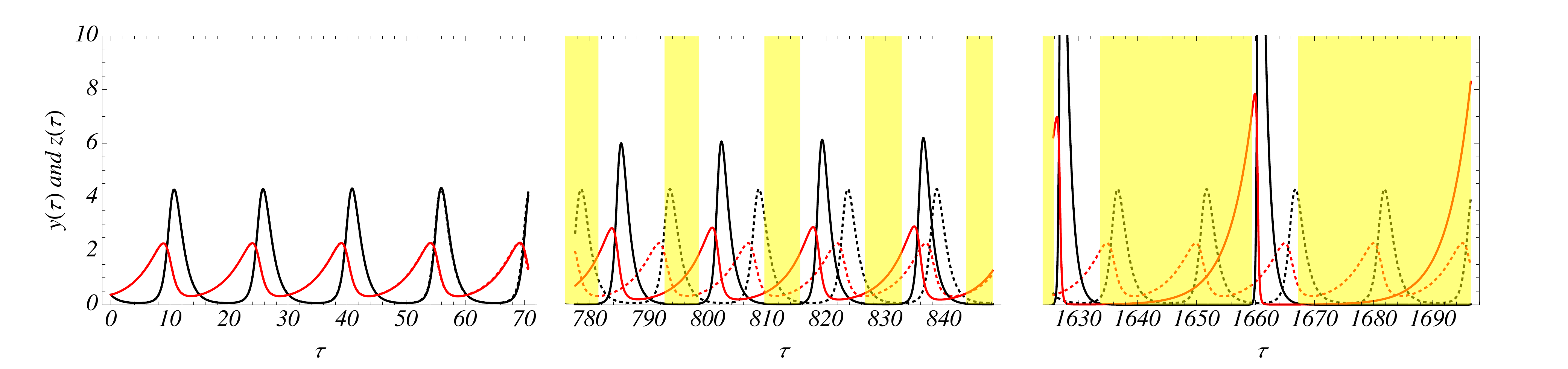}
\renewcommand{\baselinestretch}{.6}
\caption{\label{altBio04}
\footnotesize
(Color online)
Suppressed revival windows for increasing times for the same parameters from Fig.~\ref{altBio03} ($a=1/4$ and $\alpha = 1/4$).}
\end{figure}
In a virus-antivirus regime, for instance, it could result in the almost complete elimination/stabilization of both agents, for long time intervals. Phenomenologically, the question to be posed would be then how to create such a statistical environment centered at prey-predator equilibrium points in order to control the evolution dynamics.

\section{Discussion and Conclusions}

The hypothesis of quantum features affecting the dynamical evolution of biochemical and ecological microscopic systems was scrutinized for the Lotka-Volterra equation through phase-space WW framework tools.
Generalized Liouvillian and stationary properties statistically driven by gaussian ensembles were recovered \cite{Novo2021BB,Novo2023} in order to evince the quantum distortions over the classical phase-space Hamiltonian solutions.
The framework was adapted for describing the species evolution in corresponding prey-predator systems in order to evaluate the quantum imprints over them. Our results suggest that, in case of microscopic systems for which the quantum approach is relevant, the evinced instability patterns on the prey-predator phase-space trajectories can be interpreted as an output of quantum origin.
Considering that QM has already been considered for biological phenomena at lower levels of ecological organization, the non-deterministic analysis set as the background for constructing the species occurrence evolution depicted by the present framework provides a novel tool for accessing and interpreting quantum effects on such systems. 
Our results quantitatively show that the predator-prey systems of microscopical nature are sensitive to the fundamental quantum mechanical paradigms which produce measurable quantum fluctuations patterns which can be detected either directly by averaged out statistical calculations (cf. Sec.~III) or indirectly by quantum topological phase observations (cf. \cite{Novo2023}).

Given that our results are supported by a systematic Wigner flow analysis, it is our belief that they contribute to shed some light on the quantitative investigation of ecological and biochemical chains from a broader perspective where quantum-like effects can be relevant.

\vspace{.5 cm}
{\em Acknowledgments -- The work of AEB is supported by the Brazilian Agencies FAPESP (Grant No. 20/01976-5 and Grant No. 23/00392-8) and CNPq (Grant No. 301485/2022-4).}


\end{document}